\documentclass[preprint]{aastex631}

\begin{document}

\title{Measurements of the Crab Pulsar's Giant Radio Pulse Amplitude Power-Law Index Using Low-Frequency Arecibo and Green Bank Telescope Observations}

\correspondingauthor{Fronefield Crawford}
\email{fcrawfor@fandm.edu}

\author[0000-0002-2578-0360]{Fronefield Crawford}
\affiliation{Department of Physics and Astronomy, Franklin and Marshall College, P.O. Box 3003, Lancaster, PA 17604, USA}

\author{T.~Joseph~W.~Lazio}
\affiliation{Jet Propulsion Laboratory, California Institute of Technnology, 4800 Oak Grove Dr., M/S~67-201, Pasadena, \hbox{CA}  91107  USA}

\author[0000-0001-5481-7559]{Alexander McEwen}
\affiliation{Center for Gravitation, Cosmology and Astrophysics, Department of Physics, University of Wisconsin–Milwaukee, P.O. Box 413, Milwaukee, WI 53201, USA}

\author[0000-0003-1226-0793]{Julia S. Deneva} 
\affiliation{George Mason University, 4400 University Dr, Fairfax VA 22030, USA}

\author[0000-0002-4049-1882]{James~M.~Cordes}
\affiliation{Cornell Center for Astrophysics and Planetary Science, and Department of Astronomy, Cornell University, Ithaca, NY 14853, USA}

\author[0000-0002-3775-8291]{Laura Spitler}
\affiliation{Max-Planck-Institut f{\"u}r Radioastronomie (MPIfR), Auf dem H{\"u}gel 69, 53121, Bonn, Germany}

\author[0000-0002-6967-7322]{Ryan F. Trainor}
\affiliation{Department of Physics and Astronomy, Franklin and Marshall College, P.O. Box 3003, Lancaster, PA 17604, USA}



\begin{abstract}
We report two low-frequency measurements of the power-law index for the amplitudes of giant radio pulses from the Crab pulsar. The two observations were taken with the Arecibo and Green Bank radio telescopes at center frequencies of 327 MHz and 350 MHz, respectively. We find best-fit values for the differential power-law index $\beta$ (where $dN/dS \propto S^\beta$ and $S$ is pulse amplitude) of $-2.63 \pm 0.05$ and $-3.6 \pm 0.5$ from the Arecibo and Green Bank data sets, respectively. Both values are broadly consistent with other values previously measured for the Crab pulsar at low radio frequencies. These reported values may be useful in future giant pulse studies of the Crab pulsar.
\end{abstract}


\section{Introduction}

Since the discovery of the Crab pulsar as a source of individual dispersed radio pulses \citep{sr68}, it has been studied as an emitter of giant pulses (see, e.g., \citealt{l15} for a review). Giant radio pulses from pulsars have been defined as pulses having energies much greater than the mean value and having amplitudes that follow a power-law distribution \citep{rj04, gsa+21}. A departure from this behavior has been observed for the Crab pulsar by \citet{cbh+04} and \citet{mml+12}, where they saw a slight excess at very large amplitudes that might be explained by rare ``supergiant'' pulses. However, such pulses would not necessarily be expected to repeat if, for example, they were due to lensing phenomena, which is possible given the role that filaments play in producing multiple images. Subsequent observations by \citet{bc19} did not show evidence for supergiant pulses in a longer set of observations, so the observed excess may be a statistical fluke (as mentioned by \citealt{cbh+04}). Here we present two new measurements of the differential amplitude power law index $\beta$ for Crab pulsar giant pulses, where $\beta$ is defined according to $dN/dS \propto S^\beta$ and where $S$ is the pulse amplitude. The measurements were obtained from two low-frequency observations taken with the Arecibo and the Green Bank telescopes.

\section{Observations and Analysis}

The Crab pulsar was observed with the Arecibo 305-m telescope in a 5-minute diagnostic observation on 2014 April 13 (MJD 56760) as part of the Arecibo 327 MHz Drift-Scan Pulsar Survey \citep{dsm+13}. This observation used an effective bandwidth of 68.75 MHz divided into 2816 channels sampled at 81.92 $\mu$s. The Crab pulsar was also observed with the Green Bank Telescope (GBT) on 2019 October 22 (MJD 58778) as part of the Green Bank Northern Celestial Cap (GBNCC) survey \citep{slr+14}. This survey used a bandwidth of 100 MHz centered on 350 MHz that was split into 4096 channels. This was the only survey beam from the GBNCC survey that overlapped with the position of the Crab pulsar. This beam had a position offset of 0.25 degrees from the Crab pulsar position and had an integration time of two minutes. This position offset is close to (but still within) the edge of the $\sim0.3$ deg beam radius at 350 MHz for the GBT. The native sampling time of 81.92 $\mu$s of the GBNCC survey beam was increased by a factor of two in our analysis, leading to an effective sampling time of 163.84 $\mu$s. At these observing frequencies, the amount of dispersion smearing experienced by Crab pulses within the frequency channels in both observations is $\sim 0.3$ ms, which significantly exceeds the sampling times. Thus, the different effective sampling times used in the two different observations had no effect on the Crab pulse detection rate.

In both observations the data were searched blindly for single pulses at a range of dispersion measures (DMs) encompassing the Crab pulsar's DM. We used the HEIMDALL single-pulse detection package in the search \citep{b12,bbb+12}.\footnote{\url{https://sourceforge.net/projects/heimdall-astro}} All of the pulses detected by HEIMDALL that were within $0.3$ pc cm$^{-3}$ the Crab's nominal DM of 56.77 pc cm$^{-3}$ \citep{bkk+16} were retained. This DM window is consistent with the DM variability observed for the Crab's giant pulses, which does not exceed this range (e.g., \citealt{ldm+22}). We note that all of the pulses that were discarded as radio frequency interference (RFI) had DMs far from the Crab's DM value (at least 20 pc cm$^{-3}$ away), so no Crab pulses were accidentally eliminated. We measured the signal-to-noise ratio (S/N) of each detected pulse with HEIMDALL. The S/N is proportional to the pulse energy given that the (narrow) Crab pulses are temporally unresolved due to the dominance of the DM smearing, and they therefore will have similar observed widths. 

A total of 1943 and 60 Crab pulses were detected with a S/N above 6 in the Arecibo and the GBT observations, respectively. We did not make any distinction between main pulses (MPs) and interpulses (IPs) in our detected sample, even though IPs are expected to constitute a large fraction (about a third) of the giant pulses detected at these low frequencies \citep{cbh+04}. We do not expect a significant difference between these two classes of pulses at low frequencies: \cite{mat+16} showed that at 325 MHz, the average power-law index values for MPs and IPs are quite close (see their Table 3). This is supported by other studies \citep{cbh+04,hej+16,lvm+22} that have found that Crab MPs and IPs have similar properties at frequencies below a few GHz. In contrast to this, \citet{hje+15,hej+16} have shown that at high frequencies ($\ga 5$ GHz), the Crab MP and IP properties are much different and that the number of detected IPs exceeds MPs by more than an order of magnitude. 

\begin{figure}[ht!]
\begin{center}
\includegraphics[width=3.5in]{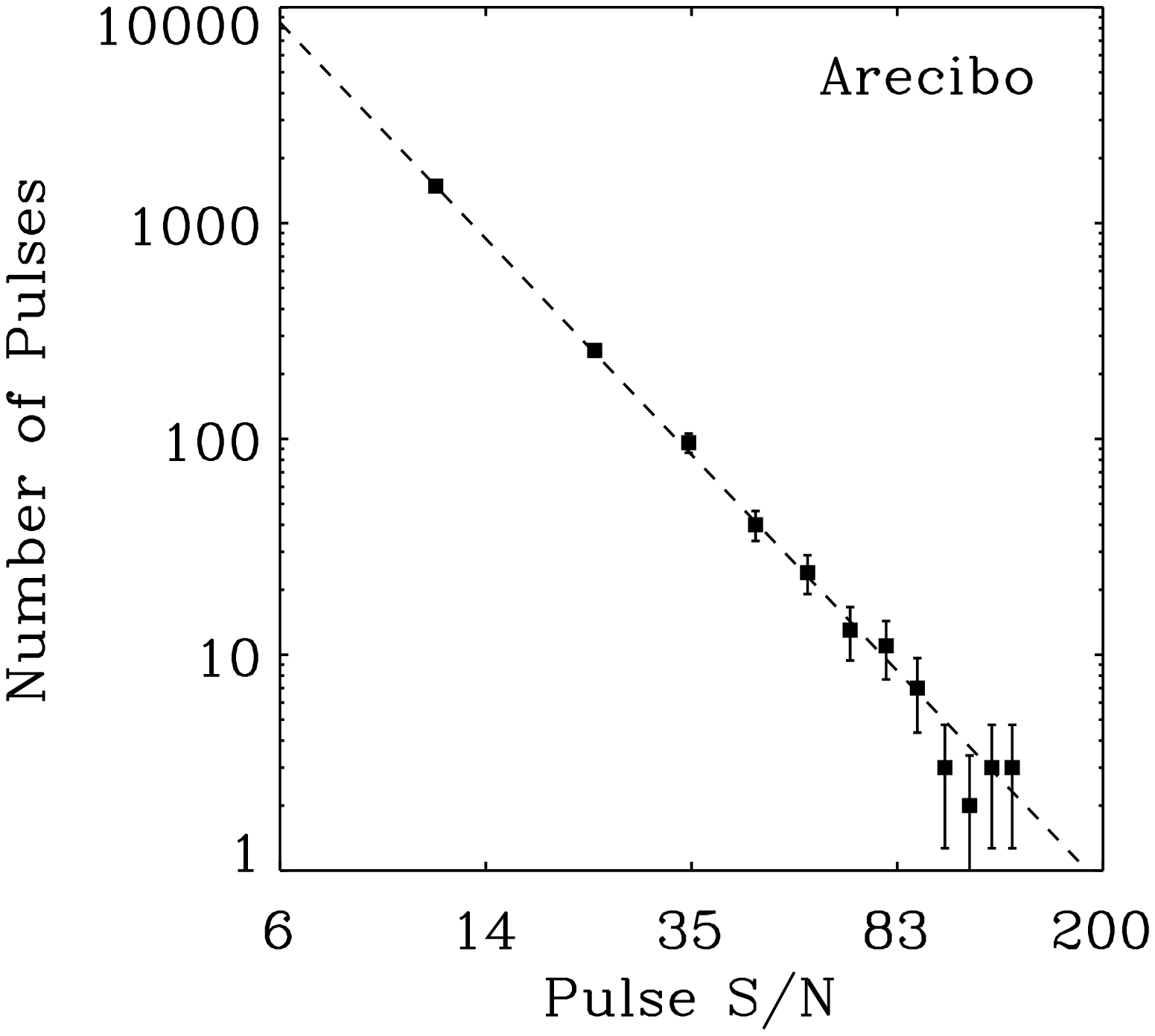}
\includegraphics[width=3.5in]{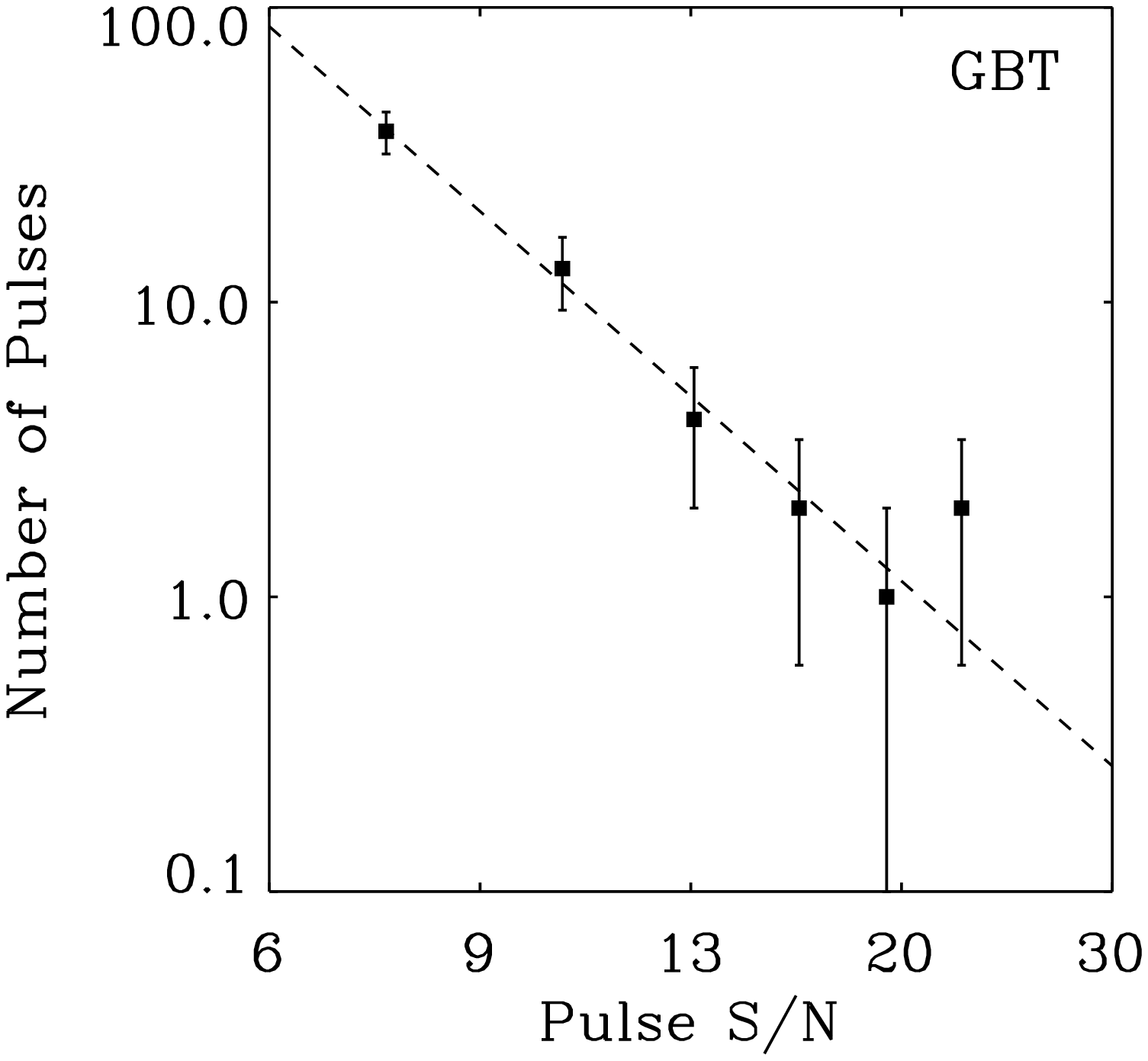}
\end{center}
\caption{Histograms of pulse amplitudes (S/N) for detected Crab pulses in observations taken with Arecibo (left) and the GBT (right). A total of 1943 pulses were detected with Arecibo in a 5-minute diagnostic observation at 327 MHz. For the GBT, 60 pulses were detected in a 2-minute observation that was part of the GBNCC survey at 350 MHz. In each case the error bars were computed as the square root of the number of pulses in each bin. The best-fit values from power law fits to the distributions are also plotted and yielded differential power-law index values of $-2.63 \pm 0.05$ and $-3.6 \pm 0.5$ for the Arecibo and GBT data sets, respectively.}
\label{fig-1}
\end{figure}

We produced histograms of the pulse amplitudes using the Sturges criterion for the binning, where the number of bins $k$ is determined by the total number of events $n$ in the histogram, according to $k = 1 + \log_{2} n$. Error bars were calculated by taking the square root of the number of events in each bin, in accordance with Poisson statistics.  Each data set was separately fit using a power law of the form $dN/dS \propto S^\beta$, where both a coefficient and a power-law index were fit as free parameters. Fig. \ref{fig-1} shows these two histograms and the best fits. 

\section{Results and Discussion}

Scaling the GBT integration from~2 to~5~minutes (the integration time of the Arecibo observation) would result in about 150 detected pulses in the GBT dataset. This number is still much smaller than the almost 2000 pulses detected with Arecibo. The telescope gain difference is not a significant factor in this difference since the Crab nebula dominates the system temperature in both observations, as shown below. The Crab nebula's flux density at 350 MHz is $S \approx 1270$ Jy (derived from the relation $S = 955 f^{-0.27}$ Jy, where $f$ is the observing frequency in GHz; \citealt{bkf+97}), while Arecibo's system equivalent flux density (SEFD) is 10 Jy and the GBT's is 35 Jy. The Crab nebula (with a characteristic diameter 5.5$'$; \citealt{cbh+04}) is also unresolved in both observations: the beam diameters of Arecibo and the GBT at these frequencies are 15$'$ and 36$'$, respectively. Thus, all of the flux from the Crab nebula was received by the telescope in each observation (see, e.g., \citealt{cbh+04}). In both cases the telescope SEFD is more than an order of magnitude smaller than the Crab nebula's flux density and does not significantly increase the system noise. Therefore, despite their difference in raw sensitivity, both Arecibo and the GBT have the same effective sensitivity to Crab giant pulses. The small difference in the central observing frequencies would also not affect the detection rates to this degree, nor would the sampling rate difference in the observations (see above). The RFI in each observation was also minimal (less than 0.1\% of the data was masked in each case). However, the two observations were conducted at epochs separated by about 5.5 years. \citet{lcu+95} showed that refractive interstellar scintillation (RISS) produces day-to-day variability for Crab giant pulses that affects the detection rate, so RISS may account for most of this difference in the detection rates.

Our best fits to the two histograms yielded differential power-law index values $\beta$ of $-2.63 \pm 0.05$ and $-3.6 \pm 0.5$ for the Arecibo and GBT data, respectively. An alternate method of estimating the power-law index using integrated number counts in a maximum likelihood estimate has been outlined by \citet{cjm70} and \citet{jem+19}. This method avoids binning of the data and is considered to be less biased. We fit our two data sets using this approach as a check, and we obtained differential power law indices (derived from the cumulative power law index) of $-2.45 \pm 0.03$ and $-3.7 \pm 0.4$ for the Arecibo and GBT data sets, respectively. These values are similar to those we obtained with binning, suggesting that both approaches produce consistent results. Our best-fit power-law values from the two data sets do not overlap within the stated (1$\sigma$) errors. This difference may be attributable to the large amount of time separating the two observations (5.5 years). Variability in the power-law index has been previously observed on long time-scales (e.g., \citealt{rps17}). 

Our power-law fit values are broadly consistent with previous low-frequency measurements from several different telescopes. \citet{oob+15} measured a power law index of $-3.35 \pm 0.35$ for the fluence distribution using the Murchison Widefield Array (MWA) at a center frequency of 193 MHz (see their Fig. 2). Observations taken with the Low Frequency Array (LOFAR) at a similarly low center frequency of 150 MHz by \citet{vmk+20} showed an amplitude index for the fluence of $-3.04 \pm 0.03$ (see their Fig. 3). From measurements at 325 MHz with the Iitate Planetary Radio Telescope (IPRT) that were taken just a few months after our Arecibo observation, \citet{mat+16} measured $\beta = -2.61^{+0.13}_{-0.15}$ for the Crab main pulse (see their Fig. 4 and Table 3). This is remarkably close (within the formal uncertainties) to the value of $-2.63 \pm 0.05$ we obtained from our Arecibo observation at essentially the same central observing frequency (327 MHz). The Crab nebula's flux density slightly exceeds the SEFD of the IPRT (\citealt{mat+16}; see their Table 3), so the IPRT system noise is comparable to the Crab nebula contribution. This makes it slightly less sensitive to Crab giant pulses compared to Arecibo, where the SEFD is negligible relative to the Crab nebula contribution (see above). The similarity of the Arecibo power-law index value to the IPRT value indicates that even though the two observations were separated by a few months, a single power-law dependence extends to a slightly lower fluence level than what was measured with the IPRT. Table 4 of \citet{mml+12} listed four previously published low-frequency measurements (taken between 150 and 430 MHz) for the Crab differential power-law index \citep{ag72,cbh+04,bwk+07,sl09} plus one new 330 MHz measurement using the Green Bank 43-m telescope. Combining these five measurements from this table yields an average power-law index of $-2.9 \pm 0.5$ for the low-frequency regime ($< 500$~MHz). Table \ref{tbl-1} shows a summary of these and other low-frequency measurements.

\begin{deluxetable}{lcll}
\tablecaption{Measured Crab Giant Pulse Power-Law Indices at Low Radio Frequencies\label{tab:spectra}}
\tablewidth{0pt}
\tablehead{
\colhead{Frequency Range} \vspace{-0.1cm} &  \colhead{Power-Law} & \colhead{Telescope} & \colhead{Reference}\\
\colhead{(MHz)} & \colhead{Index}
}
\startdata
20--84         & \ldots            & LWA1  & \cite{esd+16} \\
20--84         & \ldots            & LWA1  & \cite{ecc+13} \\
23             & \ldots            & UTR-2 & \cite{pku+06} \\
111            & \ldots            & LPA   & \cite{pku+06} \\
600            & \ldots            & RT-64 & \cite{pku+06} \\
110--180       & $-1.73 \pm 0.45$  & WSRT & \cite{rwl12} \\
185--200       & $-3.35 \pm 0.35$  & MWA & \citet{oob+15} (Figure~2) \\
111--189       & $-3.04 \pm 0.03$  & LOFAR & \citet{vmk+20} (Figure~3) \\
325            & $-2.61^{+0.13}_{-0.15}$         & IPRT & \citet{mat+16} (Figure~4 and Table~3) \\
112--430       & $-2.9 \pm 0.5$    & multiple & \citet{mml+12} (Table~4) \\
\hline
293--361       & $-2.63 \pm 0.05$  & Arecibo  & this work \\
300--400       & $-3.6 \pm 0.5$              & GBT & this work \\ 
\enddata 
\tablecomments{The table lists differential (not cumulative) power-law indices. \cite{ecc+13} do not provide an estimated power-law index due to concerns about calibration and correcting for the flux density of the Crab Nebula. \cite{esd+16} do not provide an estimated power-law index due to concerns about the small number of giant pulses detected. \cite{pku+06} do not report a power-law index. Telescope abbreviations are: LWA1 = Long Wavelength Array Station~1, UTR-2 = Ukrainian T-shaped Radio Telescope (Second Modification), LPA = Large Phased Array (Pushchino), MWA = Murchison Widefield Array, WSRT = Westerbork Synthesis Radio Telescope, RT-64 = Kalyazan 64-m Radio Telescope, LOFAR = Low Frequency Array, IPRT = Iitate Planetary Radio Telescope.}
\label{tbl-1}
\end{deluxetable}

At higher observing frequencies, various studies have measured the power-law index. \citet{mml+12} (Table 4) listed measured values ranging from $-2.1$ to $-4.1$ for frequencies between 600 and 4850 MHz. \citet{bc19} measured a value of $-2.81 \pm 0.05$ at 1330 MHz, consistent with this range. Some of this variability in values can be attributed to RISS (both \citealt{lcu+95} and \citealt{mml+12} observed day-to-day variability in the measured power law index at 812 MHz and 1.2 GHz, respectively). Note, however, that \citet{bc19} did not see variability in the power-law index on time-scales of a few days in observations taken at 1330 MHz, though on much longer time-scales (a few years) it has been observed to vary \citep{rps17}. \citet{mml+12} do not see a clear trend in how the power law index changes with frequency, but \citet{cbh+04} report a possible steepening of the index from 430 MHz to 8.8 GHz. 

In conclusion, we have measured power-law index values for the Crab giant pulse amplitude distribution using two separate low-frequency observations taken with Arecibo and the GBT. The best-fit values are broadly consistent with values previously measured at low-frequencies with different telescopes. These measurements may be useful in future giant pulse studies of the Crab pulsar.

\begin{acknowledgments}
We thank the anonymous referee for helpful comments and corrections that have improved the manuscript. The Arecibo Observatory is a facility of the National Science Foundation operated under cooperative agreement by the University of Central Florida and in alliance with Universidad Ana G. Mendez, and Yang Enterprises, Inc. The Green Bank Observatory is a facility of the National Science Foundation operated under cooperative agreement by Associated Universities, Inc. Part of this research was carried out at the Jet Propulsion Laboratory, California Institute of Technology, under a contract with the National Aeronautics and Space Administration. The NANOGrav project receives support from  National Science Foundation (NSF) Physics Frontiers Center award numbers 1430284 and 2020265. J.S.D. was supported by the National Science Foundation (AST-2009335). R.F.T is supported by the Pittsburgh Foundation (UN2021-121482) and the Research Corporation for Scientific Advancement (28289).
\end{acknowledgments}

\vspace{5mm}
\facilities{Arecibo, GBT}

\software{HEIMDALL \citep{b12,bbb+12},  
          }

\bibliography{clm+23}{}
\bibliographystyle{aasjournal}

\end{document}